\documentclass[pre,showpacs]{revtex4}  
\begin{document}
\title{Large Time Dynamics and Aging of a polymer chain in a random potential} 
\author{Yadin Y. Goldschmidt} 
\affiliation{Department of Physics and Astronomy,  
  University of Pittsburgh, Pittsburgh, Pennsylvania 15260} 
\date{\today} 
\begin{abstract} 
We study the out-of-equilibrium large time dynamics of a gaussian
polymer chain in a quenched random potential. The dynamics studied is
a simple Langevin dynamics commonly referred to as the Rouse
model. The equations for the two-time correlation and response
function are derived within the gaussian variational
approximation. In order to implement this approximation faithfully,
we employ the supersymmetric representation of the Martin-Siggia-Rose 
dynamical action. For a short
ranged correlated random potential the equations are solved
analytically in the limit of large times using certain assumptions concerning the
asymptotic behavior. Two possible dynamical behaviors are
identified depending upon the time separation- a stationary
regime and an aging regime. In the stationary regime time translation invariance
holds and so is the fluctuation dissipation theorem. The aging regime which occurs for
large time separations of the two-time correlation functions is 
characterized by history dependence and the breakdown of certain equilibrium 
relations. The large time limit of the equations yields equations
among the order parameters that are similar to the equations obtained
in the statics using replicas. In particular the aging solution
corresponds to the broken replica solution.
But there is a  difference in one equation that
leads to important consequences for the solution. The 
stationary regime corresponds to the motion of the polymer inside a 
local minimum of the random potential, whereas in the aging regime 
the polymer hops between different minima. As a byproduct we also 
solve exactly the dynamics of a chain in a random potential with 
quadratic correlations. 
\end{abstract} 
\pacs{05.40.-a, 05.70.Ln, 36.20.Ey, 75.10.Nr}
\maketitle  
 
\section{Introduction} 

The behavior of polymer chains in random media attracted much interest
in recent years because of its relevance and applications in diverse
fields \cite{chakrabarti,mut}.  Besides elucidating the properties of
the polymers chains themselves which is of much interest in physical chemistry
\cite{asher} and biology \cite{viovy}, 
this problem is directly related to the statistical mechanics
of a quantum particle in a random potential \cite{gold2}, the
behavior of flux lines in superconductors in the 
presence of columnar defects \cite{nelson,gold3}, and the problem of 
diffusion in a random catalytic environment \cite{nattermann}.
It was found in Refs.~[\onlinecite{edwards,cates,nattermann,gold,shifgold2}]
that a very long Gaussian chain, immersed in a 
random medium with very short range correlations of the disorder, will
typically curl up in some small region of
low potential energy.  The polymer chain is said to be localized and
for long chains the radius of gyration or the
end-to-end distance becomes independent of chain length ($R^2\sim L^0$).

Both heuristic arguments \cite{cates} and a variational solution of
the problem using replicas \cite{gold}, yielded the dependence of the
size of the trapped polymer on the variance of the random potential
($g$), and the logarithm of the volume of the medium (${\cal V}$) such
that $R \sim (g \ln {\cal V})^{-1/(4-d)}$ for $1\le d<4$. The breaking
of replica symmetry was crucial to the derivation of the subtle $\ln
{\cal V}$ dependence.
In a related paper \cite{gold2}, it was found that a quantum particle 
in a random environment exhibits
glassy behavior at low temperatures. The low temperature limit for a 
quantum particle translates into the long chain
limit for polymers. It implies that the free energy landscape
of a long chain is typically very complicated and possesses many 
metastable states. In a recent publication \cite{shifgold2} we further utilized this
mapping to give a physical
interpretation of the localization and glassy behavior of a polymer 
in a random potential by
making a connection with the localization of a quantum particle in a
disordered medium of finite volume. Subsequently,
we treated the case of random obstacles as opposed to a random potential,
\cite{shifgold3} and finally included the effect of a self-avoiding interaction.
\cite{shifgold4,shifgold5}. 
 
Recently methods have been developed to study analytically
the large time, non-equilibrium behavior of glassy systems. \cite{CD,CKD,KHF} 
Directed polymers and manifolds have been investigated at the mean field level,
and the solution exhibits two asymptotic time regimes: a stationary
dynamic regimes at large but similar times and a slow aging regime for
large and widely separated times. This large-time solution of the
dynamical equations has many features in
common with the replica-symmetry-breaking (RSB) solution of the
corresponding equations of the statics, although replicas are not
actually used and the limit $n \rightarrow 0 $ is avoided.
But the equations, in particular for the case of 1-step RSB,  are not
all the same in the large time limit as
the equilibrium equations. This leads to a situation that the system 
fails to reach the ultimate equilibrium state starting from an arbitrary initial
condition, and ends up in a state with higher free energy than the one
found at equilibrium. Thus the importance of the dynamical approach is
twofold: to make contact with experiments that exhibit slow relaxation and aging
effects \cite{struik} and also to serve in some cases as an alternative to the replica
approach and the $n \rightarrow 0$ limit, although this has to be
taken with a grain of salt as indicated above.

It is our goal here to extend the previous treatment of large time
dynamics for particles and
directed manifolds in quenched random environment \cite {CD,CKD} to the case of real
polymer chains. The difference between the case of directed polymers
and ``real'' polymers is mainly in the form of the random
potential. If $s$ denotes the bead number of the chain, for ``real ''
chains different beads at the same spatial locations should be exposed to the same
external potential, whereas for directed polymers different segments
always feel a different random potential even at the same transverse
position. This is easily made clear for flux lines in a
superconductors. If we have point disorder different in different x-y
planes when moving along the $z$-direction then the problem
corresponds to directed polymers in a quenched random potential. On
the other hand if we have randomly positioned columnar defects i.e. the random potential
is independent of $z$, then when projected on a single plane we see
that the system maps to ``real''  polymer chains in a quenched random
potential. In addition a real polymer might have self-avoiding
interactions among different beads, but these will not be considered
in this paper. 

In this paper we consider the Langevin dynamics of a single
gaussian chain embedded in a quenched random potential. This Langevin
dynamics is referred to in the literature as the Rouse model for a
polymer, and is the simplest dynamics \cite{doi}. For polymers in 
solutions a more realistic dynamics that reproduces more accurately
the experimental results is the Zimm model
that takes into account the effect hydrodynamic interactions effects. 
This kind of dynamics will not be considered in this paper and is a
project for future research. Our goal here is to consider the simplest
model that renders itself to an approximate analytical solution, 
of the long times dynamics. Various treatments of polymer
dynamics in a random potential have been considered before 
\cite{machta,ebert1,ebert2,vilgis1} using various approximations, 
renormalization group treatment, and/or numerical solutions of  
approximate equations of motion. These papers implicitly assume time
translation invariance (TTI) of the dynamical correlation function.
They try to extract the behavior of the center of mass diffusion
coefficient at short and large times, and their conclusions are not
always in total agreement. Only recently the possibility of two-time
dependence of the dynamical correlation function, i.e. aging
phenomena has been explored numerically by Monte Carlo simulations,
\cite {vilgis2} but the dependence on the waiting time is not
reported in detail in this paper. 

Our goal is also to make contact with the previous treatment of directed polymers in
random potential and by using similar methods the difference between
real and directed polymers will be elucidated. In addition, we derive the
equations for a random potential with general correlations, either
short or long range. The equations are derived using the so called
gaussian variational approximation which is a kind of a 
mean field treatment. This approximation was first introduced by
Feynman in his studies of superfluid helium and in the context of
random systems was first introduced in Refs.~[\onlinecite{shakhnovich,MP1}]. 
In order to implement it faithfully such as to preserve correctly all 
the symmetries of the dynamical equations we use the supersymmetric
(SUSY) formulation of the dynamics \cite{kurchan}, implement the 
variational approximation, and then disentangle SUSY to produce
coupled integro-differential
equations for the correlation and response functions. Here we follow
steps similar to those used by Cugliandolo et al. \cite{CD,CKD} and Konkoli 
et al. \cite{KHF} in their treatment of directed manifolds and
random heteropolymers respectively. But because the random potential
is implemented differently in our case, i.e. different beads at the
same spatial position feel the same potential, the resulting equations are
different.

In order to solve the equations analytically at large times we make some
assumptions about the limiting behavior of the correlation and
response functions at large times. These function depend on two
different times. When these times are large but their separation is
small compare to the individual times, the behavior depends only on
the separation and thus TTI holds as well
as the fluctuation dissipation theorem (FDT). On the other hand when
the separation of times becomes very large TTI and FDT break down
although a generalized form of FDT still hold. The breakdown of TTI
and FDT is one of the main the characteristics of the glassy phase and
is referred to in the literature as ``aging''. 
Although the dynamical equations are valid for random potential with quite
general correlations, in this paper we solve the equations for the
case of short ranged correlations of the random potential. This case
includes the case of $\delta$-correlated potential i.e. the potential
at different points in space are uncorrelated, but we include the case
of a short correlation length and the case that there is power law
decay of the correlation with large enough power. In this case the 
equilibrium solution involved a 1-step RSB as found in
Ref.~[\onlinecite{gold}]. We also consider in an Appendix the case of long
ranged quadratic correlation of the random potential which is exactly
soluble without the variational approximation. The case of the
solution for other long ranged correlated potential will be considered
elsewhere. For the statics this case involves continuous RSB \cite{gold2},
and thus the dynamical ansatz should be different.

\section{The Model} 
 
The model is defined as follows. The Langevin dynamics is assumed to 
be governed by the Hamiltonian $H[x]$, 
\begin{equation} 
  \partial x(s,t)/\partial t = - \partial H[x] / \partial x(s,t) + \eta(s,t), 
  \label{eq:dxdt} 
\end{equation} 
where $x(s,t)$ is a $d$-dimensional vector representing the position 
of chain bead $s$ at time $t$. Beads 
are numbered continuously from $s=0$ to $s=L$. $\eta(s,t)$ is Gaussian 
noise: 
\begin{equation} 
  \langle \eta(s,t)\eta(s',t') \rangle_T = 2\delta(s-s')\delta(t-t') T 
  \label{eq:etas} 
\end{equation} 
due to contact with a heat bath at temperature $T$. This dynamics is
the simplest dynamics for a polymer chain, referred to in the
literature as the Rouse model \cite{doi}.
The Hamiltonian $H[x]=H_0[x]+H_{rand}[x]$ contains a deterministic 
part $H_0[x]$ and a random part $H_{rand}[x]$. The $H_0[x]$ is defined 
as 
\begin{equation} 
   H_0[x]= \frac{1}{2} \int_{0}^{L} ds  
        [M (\partial x(s,t)/\partial s)^2+\mu x(s,t)^2] 
   \label{eq:H0} 
\end{equation} 
where $M =d\ T /b_K^2$ , d is the number of spatial dimensions and
$b_K$ is the Kuhn bond length of the polymer. This representation is
the simplest representation of a polymer as a gaussian chain in the
continuum approximation. The parameter  
$\mu$ plays the role of a finite volume since the polymer is
confined by the harmonic potential to a finite region of space. Thus $\mu\rightarrow 0$ is the
large volume limit and $|\ln \mu| \propto \ln {\cal V}$ for a volume
$\cal V$. \cite{gold} The random part 
$H_{rand}$ describes the interaction between each bead and the
external random potential: 
\begin{equation} 
   H_{rand}[x]= \int_{0}^{L} ds V(x(s,t)). 
   \label{eq:Hrand} 
\end{equation}  
$V(x)$ is a short-range potential, and for simplicity we 
take it to have a Gaussian form, 
\begin{equation} 
   \langle V(x)V(x')\rangle=\frac{g}{(d\pi\sigma^2)^{d/2}}\exp\left({-\frac{(x-x')^2}{d\sigma^2}}\right). 
   \label{eq:V} 
\end{equation} %
$d$ is the dimensionality of the system, and $\sigma$ parameterizes 
the range of the potential. In particular, for $\sigma\rightarrow 0$, 
$\langle V(x)V(x')\rangle\rightarrow g\delta(x- x')$, and we recover the potential used in \cite{gold}.
More generally we can take
\begin{equation} 
   \langle V(x)V(x')\rangle=-d J\left(\frac{(x-x')^2}{d}\right), 
   \label{eq:VG} 
\end{equation}  
for some function J(z). For the case represented by Eq.(\ref{eq:V}),
\begin{equation}
  \label{eq:Js}
  J(z)=-\frac{g}{d(d\pi\sigma^2)^{d/2}}\exp(-z/\sigma^2).
\end{equation}
This model admits a stationary solution characterized by a Gibbs 
distribution.  The equilibrium partition function for this solution is 
given by 
\begin{equation} 
  {\cal Z} = \int Dx  
      e^{ - \frac{1}{2T} \int_{0}^{L} ds  
           [M( \partial x(s)/\partial s)^2+\mu x(s)^2 + V(x(s)] }. 
  \label{eq:Z} 
\end{equation}

\section{Mapping to the Field Theory} 
 
Here we will follow closely the notation of Ref.~[\onlinecite{KHF}]. 
Using the standard Martin-Siggia-Rose formalism \cite{MSR}, the 
dynamical average of any observable can be calculated as 
\begin{equation} 
  \langle {\cal O}[x,\tilde x] \rangle_T =\int Dx D\tilde x  D\xi D\bar\xi  
  {\cal O}(x,\tilde x) e^{-S[x,\tilde x,\xi,\bar\xi] }, 
  \label{eq:average} 
\end{equation} 
with the following dynamical action: 
\begin{eqnarray} 
  S[x,\tilde x,\xi,\bar\xi] = && 
     \int dt ds 
     \left[ 
          - T \tilde x(s,t)^2  
          + \tilde x(s,t) \left(  
              \frac{\partial}{\partial t} x(s,t)  
              + \frac{\partial H[x]}{\partial x(s,t)} 
            \right)  
     \right] \cr  
     && - \int dt ds \bar\xi(s,t)\frac{\partial}{\partial t}\xi(s,t) 
     + \int dt ds ds' \bar\xi(s,t) 
         \frac{\partial^2 H[x]}{\partial x(s,t) \partial x(s',t)} \xi(s',t) 
  \label{eq:S}  
\end{eqnarray}  
$\tilde x$, $\xi$, $\bar\xi$ are auxiliary fields which appear in the 
formalism.  To make for a more compact notation we introduce the superfield $\Phi$: 
\begin{equation} 
  \Phi(s,t_1,\theta_1,\bar\theta_1)=x(s,t_1)+\bar\xi(s,t_1)\theta_1  
  + \bar\theta_1\xi(s,t_1) + \bar\theta_1\theta_1\tilde x(s,t_1), 
  \label{eq:Phi} 
\end{equation} 
where $\theta$ and $\bar\theta$ are Grassmann variables
(anti-commuting c-numbers) . For $X,X' \in \{\theta, \bar\theta,\theta',\bar\theta'\}$, 
$\{X,X'\}=0$ and $\int dX X=1$, the rest of the integrals being 
zero. In the following, for practical reasons, the more compact 
notation $\Phi(s,1)\equiv\Phi(s,t_1,\theta_1,\bar\theta_1)$ will be 
used. Also, the integral symbol $\int d\theta_1 d\bar\theta_1 dt_1$ 
will be denoted by $\int d1$. 
 
In supersymmetric (SUSY) notation Eqs.~(\ref{eq:average}) and (\ref{eq:S})  
translate into (\ref{eq:avSUSY}) and (\ref{eq:SSUSY}): 
\begin{eqnarray} 
  && \langle {\cal O}[\Phi] \rangle_T =\int D\Phi  
    {\cal O}[\Phi] e^{-S[\Phi] },           \label{eq:avSUSY} \\ 
  && S[\Phi] = S_0[\Phi]+S_{rand}[\Phi],      \label{eq:SSUSY}  
\end{eqnarray}  
where 
\begin{eqnarray} 
  && S_0[\Phi]= (1/2) \int ds d1 ds' d2 \Phi(s,1) K_{12}^{ss'} \Phi(s'2), 
  \label{eq:S0} \\ 
  && S_{rand}[\Phi]= \int d1 ds V(\Phi(s,1)), 
  \label{eq:Srand} 
\end{eqnarray} 
and 
\begin{eqnarray} 
  K_{12}^{ss'} && \equiv \delta_{12} \delta_{ss'} K_1^s \ , \ \ 
  K_1^s  = T \left[ \mu/T-(\partial/\partial s)^2 \right] - D_1^{(2)}, \\  
  D_1^{(2)} && =2 T \frac{\partial^2}{\partial\theta_1\partial\bar\theta_1} + 
   2 \theta_1 \frac{\partial^2}{\partial\theta_1\partial t_1} -  
  \frac{\partial}{\partial t_1}, 
\end{eqnarray} 
As noticed by De Dominicis \cite{Dom} the expression in 
Eq.(\ref{eq:avSUSY}) is already normalized, so the average over the 
quenched random interactions $V$ can be done directly on 
(\ref{eq:avSUSY}): 
\begin{equation} 
  \langle\langle A[\Phi] \rangle_T\rangle_V = \int D\Phi 
  A[\Phi] e^{-(S_0[\Phi]+S_1[\Phi])}, 
  \label{eq:avA} 
\end{equation} 
where $\exp(-S_1[\Phi])\equiv\langle\exp(-S_{rand}[\Phi])\rangle_V$. 
The average over $V$ can be done easily, leading to 
\begin{equation} 
  S_1[\Phi] = \frac{d}{2}\int ds ds' d1 d2 J\left(\frac{(\Phi(s,1)-\Phi(s',2))^2}{d}\right).
\end{equation} 
The dynamical action $S_{AV}=S_0+S_1$ closely resembles the effective 
Hamiltonian obtained in the static replica approach of 
refs. \cite{gold2,gold}.  This rather general similarity between replica 
and SUSY treatments has been discussed in ref. \cite{kurchan}.  Instead 
of summation over replica indices in \cite{gold2,gold} we have $\int 
d1 \int d2$.

\section{Variational approximation} 
 
Since the model cannot be solved exactly, we proceed by using a
variational approximation, first introduced by Feynman. We Assume 
that fields $\Phi$ are approximately described by a quadratic action 
\begin{equation} 
  S_{var}=\frac{1}{2} \int d1 ds d2 ds'  
    \Phi(s,1) G(s,1;s',2)^{-1} \Phi(s',2).  
\end{equation} 
This approach has been widely used in statics.  Here we apply it to a 
dynamic calculation.  The goal is to calculate $F_{dyn}$ , given
formally by
\begin{equation} 
  e^{-F_{dyn}} = e^{-\langle(S_{AV}-S_{var})\rangle_{var}} e^{-f_{var}} 
  \label{eq:FJvar}, 
\end{equation} 
where 
\begin{equation} 
  e^{-f_{var}}=\int D\Phi e^{-S_{var}} \ ,\ \ \ \  \langle 
  . \rangle_{var}=e^{f_{var}}\int D\Phi ( . ) e^{-S_{var}}. 
\end{equation} 
In usual statics, for problems without disorder, the variational 
approach is related to a maximum principle.  The equivalent of 
Eq.(\ref{eq:FJvar}) leads to the inequality 
\begin{equation} 
   e^{-F} \ge e^{  - \langle (S_{AV}-S_{var}) \rangle_{var} } 
              e^{ -f_{var} }. 
   \label{eq:FJFdyn} 
\end{equation} 
In the present dynamical problem, as well as in the static problem 
with replicas, such a maximum principle is not known, 
and the variational free-energy cannot be claimed to be an upper bound 
to the true one. Despite that, the variational approach has been 
argued to give exact results for directed manifolds in the limit of infinite embedding
dimensions \cite{MP1,MP2}, giving some justification for its use
even at finite dimensions. For real polymers we obtain a meaningful
solution for $1 \le d < 4$ and we cannot use the large $d$ limit directly.
 
The dynamical variational free-energy $F_{dyn}=\langle (S_{AV}-S_{var}) 
\rangle_{var}+f_{var} $ is given by 
\begin{equation} 
F_{dyn} = F_{dyn}^{(1)} + F_{dyn}^{(2)} + F_{dyn}^{(3)}, 
\label{eq:Fdyn} 
\end{equation} 
with 
\begin{eqnarray} 
  && F_{dyn}^{(1)} = \frac{d}{2} \int ds d1 ds' d2  
     K_{12}^{ss'} G_{12}^{ss'}  \label{eq:Fdyn1} \\  
  && F_{dyn}^{(2)} = - \frac{d}{2} Tr \ln G \label{eq:Fdyn2} \\ 
  && F_{dyn}^{(3)} = \frac{d}{2}\int ds d1 ds' d2 \left\langle 
J\left(\frac{(\Phi(s,1)-\Phi(s',2))^2}{d}\right)\right\rangle_{var}.  \label{eq:Fdyn3} 
\end{eqnarray} 
We proceed to calculate the last term $F^{(3)}_{dyn}$. Using the identity 
 
\begin{eqnarray}
  \left\langle J((\Phi-\Phi')^2/d)\right\rangle_{var}=\int d^dy J(y^2/d)\int
  \frac{d^dp}{(2\pi)^d}\exp(-ip \cdot y)\left\langle \exp(i p \cdot (\Phi-\Phi')\right\rangle_{var},
\label{eq:Jvar}
\end{eqnarray}
it is easy to verify that
\begin{eqnarray}
  \label{eq:B}
 \left\langle \exp(i p \cdot
   (\Phi-\Phi')\right\rangle_{var}=\exp(-\frac{1}{2}p^2 B_{12}^{ss'}), 
\end{eqnarray}
where 
\begin{eqnarray}
  B_{12}^{ss'}=G(s,1;s,1)+G(s',2;s',2)-2G(s,1;s',2). \label{eq:BG}
\end{eqnarray}
Defining
\begin{eqnarray}
\hat{J}(a)  & \equiv\int d^d{ y\ }J({ y}^{2}/d)\int
\frac{d^d{ p}}{(2\pi)^{d}}\exp(-i{  p\cdot y})\exp\left(
-\frac{a{ p}^{2}}{2}\right) \nonumber\\
\  & =\frac{1}{\Gamma(d/2)}\int_{0}^{\infty}dx\ x^{d/2-1}e^{-x}J\ \left(
\frac{2xa}{d}\right)  ,\label{eq:Jhat}%
\end{eqnarray}
we observe, by substituting Eq.(\ref{eq:B}) in Eq.(\ref{eq:Jvar}) that
\begin{eqnarray}
\left\langle J((\Phi-\Phi')^2/d)\right\rangle_{var}=\hat{J}(B_{12}^{ss'}).  
\end{eqnarray}
Thus
\begin{eqnarray}
\label{eq:Fdyn3V}
  F_{dyn}^{(3)} = \frac{d}{2}\int ds d1 ds' d2 \hat{J}(B_{12}^{ss'}).
\end{eqnarray}
For the case that $J$ is given by Eq.(\ref{eq:Js}) we find
\begin{eqnarray}
  \hat{J}(a)=-\frac{g}{d(2\pi)^{d/2}}\left(\frac{d\sigma^2}{2}+a\right)^{-d/2},
\label{Jhatgauss}
\end{eqnarray}
and recall that $\sigma \rightarrow 0$ for a $\delta$-correlated potential.
If $J(a)$ is of the form
\begin{equation} 
  J(a)= \frac{g a^{1-\gamma}}{2(1-\gamma)},
  \label{eq:Jpower} 
\end{equation}
for large $a$, i.e. it involves power law correlations of the disorder, then
\begin{equation} 
  \hat{J}(a)= \frac{\hat{g} a^{1-\hat{\gamma}}}{2(1-\hat{\gamma})}, 
  \label{eq:Jhatpower} 
\end{equation}  
for large $a$, where \cite{MP1} 
\begin{eqnarray}
\hat{\gamma}&=&\gamma \ \ \ \ \ \ \ \ \ \   {\rm if} \ \ \ \gamma<1+d/2 \\
\hat{\gamma}&=&1+d/2\ \ {\rm if} \ \ \  \gamma \geq 1+d/2\\
\hat{g}&\propto& g.
\end{eqnarray}

\section{Equations of motion in supersymmetric notation}

Given the $F_{dyn}$,  one can derive the equations of motion 
from the stationarity condition 
\begin{equation} 
  \frac{\delta}{\delta G_{12}^{ss'}} F_{dyn} = 0.  
  \label{eq:dFdyn} 
\end{equation} 
The most complicated term  is $\frac{\delta}{\delta 
G_{12}^{ss'}} F_{dyn}^{(3)}$.  From (\ref{eq:Fdyn3V}), it is   
\begin{equation} 
  \frac{d}{2} 
  \int d3 d4 du dv  
     {\hat J}\ ' \left[ B_{34}^{uv}\right]  
     (\delta_{ss'}\delta_{us}\delta_{13}\delta_{23} + \delta_{ss'}\delta_{vs}\delta_{14}\delta_{24} -  
       \delta_{us}\delta_{vs'}\delta_{13}\delta_{24} - \delta_{us'}\delta_{vs}\delta_{14}\delta_{23} ). 
  \label{eq:dFdyn3V} 
\end{equation} 
Eq.(\ref{eq:dFdyn3V}) simplifies to 
\begin{equation} 
    \frac{\delta}{\delta G_{12}^{ss'}} F_{dyn}^{(3)} =  
    d \, \left[\delta_{ss'} \delta_{12} \int d3 du {\hat{J}}'(B_{13}^{su}) - {\hat{J}}'(B_{12}^{ss'})  
      \right]. 
\end{equation} 
The variations of $F_{dyn}^{(1)}$ and $F_{dyn}^{(2)}$ are trivial. 
Using Eq.~(\ref{eq:dFdyn}) and (\ref{eq:Fdyn}) leads to 
\begin{equation} 
  K_{12}^{ss'} - ( G_{12}^{ss'} )^{-1} 
  + 2 \, \left[\delta_{ss'}   
        \delta_{12} \int d3du {\hat{J}}'(B_{13}^{su}) - {\hat{J}}'(B_{12}^{ss'})  
    \right] = 0, 
  \label{eq:KGinv} 
\end{equation} 
which can be written as 
\begin{equation} 
  K_1^s G_{12}^{ss'} = \delta_{12} \delta_{ss'}  
   + 2 \int d3du {\hat{J}}'(B_{13}^{su})(G_{32}^{us'}-G_{12}^{ss'}). 
  \label{eq:emSUSY} 
\end{equation} 
Due to translational invariance in the variable $s$, $G$ depends only
on the difference $s-s'$. Thus 
\begin{eqnarray}
  G_{12}^{ss'}=G_{12}^{s-s'}.
\end{eqnarray}
It is useful to define following Fourier transforms 
\begin{equation} 
  G_{12}^{s} \equiv \frac{1}{L}\sum_k 
  e^{-iks}\hat{G}_{12}^k. 
\end{equation} 
Since $0<s<L$ the corresponding wave numbers $k$ take the values
$k=(2\pi/L)n$ where $n=0,\pm 1,\pm2,\cdots$. In the following it will
become necessary to separate the $k=0$ component from $k\neq 0$.
Also
\begin{eqnarray}
  \label{eq:invft}
\hat{G}_{12}^k=\int_0^L ds e^{iks}G_{12}^s.
\end{eqnarray}
Then Eq.(\ref{eq:emSUSY}) translates into 
\begin{equation} 
   \left[ \mu+T k^2-D^{(2)}_1 \right] \hat{G}_{12}^k = \delta_{12} 
    + 2 \int d3du
    {\hat{J}}'(B_{13}^{u})(e^{iku}\hat{G}_{32}^k-\hat{G}_{12}^k),  
  \label{eq:emSUSYk} 
\end{equation}
 where
 \begin{eqnarray}
   B_{13}^u=\frac{1}{L}\sum_{k'} (\hat{G}_{11}^{k'}+\hat{G}_{33}^{k'}-2e^{-ik'u}\hat{G}_{13}^{k'}).
 \end{eqnarray}

\section{Disentangling supersymmetry}

$G_{12}^{ss'}$ encodes 16 correlation functions, out of which only 
two, correlation and response function, are independent and nonzero:  
\begin{eqnarray} 
  && \langle \langle x(s,t_1) x(s',t_2) \rangle \rangle/d  
     \equiv C(s,t_1;s',t_2) = 
     \frac{1}{L}\sum_k e^{ik(s-s')}C_k(t_1,t_2) \label{eq:C} \\ 
  && \langle \langle x(s,t_1) \tilde x(s',t_2) \rangle \rangle /d 
     \equiv R(s,t_1;s',t_2) = 
     \frac{1}{L}\sum_k  e^{ik(s-s')} R_k(t_1,t_2). \label{eq:R} 
\end{eqnarray} 
Also, by adding an external field term to the original Hamiltonian $H[x] 
\rightarrow H[x]+\int ds dt x(s,t) h(s,t)$ one gets 
\begin{equation} 
  \langle\langle x(s,t_1) \tilde x(s',t_2) \rangle\rangle = 
   \frac{\delta}{\delta h(s',t_2)} \langle\langle x(s,t_1)\rangle\rangle. 
\end{equation} 
i.e. $R(s,t_1;s',t_2)$ describes the response to an infinitesimal 
field applied at time $t_2$ and bead $s'$. Thus, $G_{12}^k$ reduces to 
\begin{equation} 
  G_{12}^k = C_k(t_1,t_2) +  
    (\bar\theta_1-\bar\theta_2) \left[ \theta_1 R_k(t_2,t_1) -  
    \theta_2 R_k(t_1,t_2) \right]. 
  \label{eq:G12k} 
\end{equation} 
It follows that 
\begin{equation}
\label{eq:Bu12} 
  B_{12}^u = B^u(t_1,t_2) - \frac{2}{L}\sum_k e^{-ik'u}(\bar\theta_1-\bar\theta_2)  
    \left[ \theta_1 R_{k'}(t_2,t_1) - \theta_2 R_{k'}(t_1,t_2) \right], 
\end{equation} 
with 
\begin{equation} 
\label{eq:Bu}
   B^u(t_1,t_2)\equiv  \langle \langle (x(u,t_1)- x(0,t_2))^2\rangle \rangle/d = \frac{1}{L}\sum_k  
                    \left[ 
                       C_k(t_1,t_1)+C_k(t_2,t_2)-2 e^{-iku}C_k(t_1,t_2) 
                    \right]. 
\end{equation} 
After disentangling the equations of motion in SUSY notation (see 
Eq.~\ref{eq:emSUSYk}) by using (\ref{eq:G12k}-\ref{eq:Bu}) gives  
\begin{eqnarray} 
 & [\mu+Mk^2+\partial/\partial t] C_k(t,t') = &  
  2 T R_k(t',t) + 
  2 \int_{0}^{t'} dt_3 \int_0^Ldu {\hat{J}}'\left[B^u(t,t_3)\right] R_k(t',t_3)e^{iku} + \nonumber \\ 
  & & + 4 \int_{0}^{t} dt_3 \int_0^L du {\hat{J}}''\left[B^u(t,t_3)\right] R^u(t,t_3) \left[  
  C_k(t,t')-e^{iku}C_k(t_3,t') \right], \label{eq:emC} \\ 
  & [\mu+Mk^2+\partial/\partial t] R_k(t,t') = & \delta(t-t') + 
  4 \int_{0}^{t} dt_3 \int_0^L du  {\hat{J}}''\left[B^u(t,t_3)\right] R^u(t,t_3) \left[  
    R_k(t,t')-e^{iku}R_k(t_3,t') \right], \label{eq:emR} 
\end{eqnarray} 
where we defined
\begin{eqnarray}
  R^u(t,t')=\frac{1}{L}\sum_k e^{-iku} R_k(t,t').
\end{eqnarray}

\section{Ansatz for the correlation and response functions} 
\label{sec:ansatz} 
 
These equations of motion are coupled integro-differential equations 
which in principle can be solved; the initial conditions are given by 
$C_k(0,0)$ and we use Ito's convention $R(t+\epsilon,t)\rightarrow 1$ 
as $\epsilon\rightarrow 0$ from above. It is well known that asymptotic solutions 
of such equations can be characterized by few parameters and it is 
possible to solve those equations 
analytically. \cite{CD,CKD,FM,CK1,BCKP,CK2} 
 
For $t,t'\rightarrow\infty$, $\tau/t' \ll 1$ and $\tau=t-t'$, TTI holds 
\begin{eqnarray} 
  && \lim_{t\rightarrow\infty} C_k(t,t) = \tilde q_k,      \\ 
  &&  \lim_{t\rightarrow\infty} C_k(t+\tau,t) = C_k(\tau), \\ 
  && \lim_{\tau\rightarrow\infty} C_k(\tau) = q_k,   
\end{eqnarray} 
and 
\begin{equation} 
  \lim_{t\rightarrow\infty} R_k(t+\tau,t) = R_k(\tau) . 
\end{equation} 
We will refer to this regime as the stationary or TTI regime.
In addition to the TTI regime, there is another long time non trivial 
regime, characterized by $t,t'\rightarrow\infty$, fixing 
$\alpha=h(t')/h(t)$ and $0<\alpha<1$, where the function $h(t)$ is 
an increasing function of $t$ which the asymptotic analysis performed 
here is not able to determine. 
In this aging regime one has 
\begin{eqnarray} 
  && \lim_{t\rightarrow\infty} C_k(t,h^{-1} 
       (\alpha h(t))) = q_k \hat C_k(\alpha), \\ 
  && \lim_{\alpha\rightarrow 0} q_k \hat C_k(\alpha) = q_{0,k}, \\ 
  && \lim_{\alpha\rightarrow 1} \hat C_k(\alpha) = 1, 
\end{eqnarray} 
and 
\begin{equation} 
  \lim_{t\rightarrow\infty} R_k(t,\alpha t) = \frac{1}{t}\hat R_k(\alpha). 
\end{equation} 
Also, for future convenience, 
it is useful to introduce the following order 
parameters: 

\begin{eqnarray} 
  \tilde{b}(u) &=&  \frac{2}{L}\sum_k (1-e^{-iku})\tilde{q}_k , \\
   {b}(u) &=&  \frac{2}{L}\sum_k (\tilde{q}_k-e^{-iku}{q}_k ),\\
   {b_0}(u) &=&  \frac{2}{L}\sum_k (\tilde{q}_k-e^{-iku}{q}_{0,k} ).     
\end{eqnarray} 
 Also 
 \begin{eqnarray}
   B^u(\tau) &=&  \frac{2}{L}\sum_k [\tilde{q}_k-e^{-iku}C_k(\tau)],\\
  \hat{B}^u(\alpha) &=& \frac{2}{L}\sum_k [\tilde{q}_k-e^{-iku}q_k
  \hat{C}_k(\alpha)],\\
  \hat{R}^u(\alpha) &=& \frac{1}{L}\sum_k e^{-iku}\hat{R}_k(\alpha). 
 \end{eqnarray}

\section{Equations relating asymptotic values of correlation  
and response functions} 
\label{sec:qs} 
 
Using the ansatz discussed in section \ref{sec:ansatz} one can derive 
the following equations for $C_k(t,t')$ in the TTI regime: 
\begin{eqnarray} 
 & &  \left( \mu+T k^2+\partial/\partial\tau \right)  C_k(\tau) =
   \nonumber \\  
 & & 2 T R_k(-\tau)  
  + \frac{2}{T} \int_0^L du {\hat{J}}'[b(u)] \left[ C_k(\tau) -
    e^{iku}q_k \right] \nonumber \\
 & &- \frac{2}{T} \int_0^L du {\hat{J}}'[\tilde{b}(u)] (1-e^{iku})C_k(\tau)  
  - \frac{2}{T} \int_{0}^{\tau} d\tau' \int_0^L du e^{iku} {\hat{J}}'[B^u(\tau-\tau')]  
        \frac{\partial C_k(\tau')}{\partial\tau'} 
 \nonumber \\ 
 & &  +  2 \int_{0}^{1} d\rho \int_0^L du e^{iku}{\hat{J}}'[\hat B^u(\rho)] \hat R_k(\rho)  
      +  4 \int_{0}^{1} d\rho \int_0^L du {\hat{J}}''[\hat B^u(\rho)] \hat R^u(\rho)  
        \left[ C_k(\tau) - e^{iku}q_k \hat C_k(\rho) \right]  
  \label{eq:CkTTI} 
\end{eqnarray} 
It is also possible to derive similar equations for $R_k(\tau)$ which, 
due to the Fluctuation Dissipation Theorem (FDT) 
\begin{equation} 
  R_k(\tau) = - \frac{1}{T} \frac{d\,C_k(\tau)}{d\,\tau} 
  \label{eq:FDT}, 
\end{equation} 
are completely equivalent to Eq.~(\ref{eq:CkTTI}). 
 
In the aging regime one gets the following equation for $q_k\hat 
C(\alpha)$: 
\begin{eqnarray} 
   \left[ \mu+Mk^2 \right.   
    \left. -   4 \int_{0}^{1} d\rho \int_0^L du e^{iku}{\hat{J}}''(\hat B^u(\rho)) \hat R^u(\rho)  
    \right]  q_k \hat C_k(\alpha)  =   
    2 \int_{0}^{1} d\rho \int_0^L du e^{iku}{\hat{J}}'(\hat
    B^u(\alpha\rho)) \hat R_k(\rho)
    \nonumber \\  
    +  \frac{2}{T} \int_0^L du e^{iku}{\hat{J}}'(\hat B^u(\alpha)) ( \tilde q_k - q_k )  
   -  4 \int_{0}^{\alpha} d\rho \int_0^L du e^{iku}{\hat{J}}''(\hat B^u(\rho)) \hat R^u(\rho)     
        q_k \hat C_k(\rho/\alpha) \nonumber \\ 
     -  4 \int_{\alpha}^{1} d\rho \int_0^L du e^{iku}{\hat{J}}''(\hat B^u(\rho)) \hat R^u(\rho)     
        q_k \hat C_k(\alpha/\rho) 
   + 4 \int_0^\infty d\tau' \int_0^L du
  \hat{J}''[B^u(\tau')]R^u(\tau')(1-e^{iku})q_k \hat{C}_k(\alpha).  
  \label{eq:CkAG} 
\end{eqnarray} 
For $\hat R_k(\alpha)$ we obtain, 
\begin{eqnarray} 
   \left[  
      \mu+Mk^2  
      -  4 \int_{0}^{1} d\rho \int_0^L du {\hat{J}}''(\hat B^u(\rho)) \hat R^u(\rho)  
   \right] \hat R_k(\alpha)  = 
    - \frac{4}{T} \int_0^L du e^{iku}{\hat{J}}''(\hat B(\alpha)) \hat R^u(\alpha)  
      ( \tilde q_k - q_k ) \nonumber \\ 
     -  4 \int_{\alpha}^{1} \frac{d\rho}{\rho} \int_0^L du e^{iku}{\hat{J}}''(\hat B^u(\rho))     
        \hat R^u(\rho) \hat R_k(\alpha/\rho) + 4 \int_0^\infty
        d\tau'\int_0^L du (1-e^{iku})\hat{J}''[B^u(\tau')]R^u(\tau')\hat{R}_k(\alpha). 
  \label{eq:RkAG} 
\end{eqnarray} 
Again, one can see that both Eq.~(\ref{eq:CkAG}) and 
Eq.~(\ref{eq:RkAG}) can be solved by the ansatz 
\begin{equation} 
  \hat R_k(\alpha) = \frac{x_c}{T} q_k \frac{d\,\hat C_k(\alpha)}{d\,\alpha}.  
  \label{eq:GFDT} 
\end{equation} 
Eq.~(\ref{eq:GFDT}) is commonly referred to as a generalized FDT 
(GFDT). The parameter $x_c$ corresponds to the corresponding parameter in the
static replica solution. In the context of replicas it was first introduced by
Parisi and should not be confused with a spatial coordinate. This
parameter must satisfy the inequality $x_c<1$. In
principle, Eq.~(\ref{eq:GFDT}) could have been written as 
\begin{equation} 
  \hat R_k(\alpha) = \frac{ x_{ck}(q_k\hat C_k(\alpha)) }{ T } \,  
                     q_k \frac{ d\hat C_k(\alpha) }{ d\alpha }, 
\end{equation} 
which could be applied to a many-step RSB scheme. However, as
for the case of directed polymer with short ranged correlated random potential
we found \cite{gold} that a solution with one step RSB is appropriate, and it is sufficient to 
use the simpler ansatz given in Eq.~(\ref{eq:GFDT}). 
 
For $t=t'$ and $t\rightarrow\infty$ Eq.(\ref{eq:emC}) gives 
\begin{eqnarray} 
  (\mu+Mk^2) \tilde q_k =  
     T + \frac{2}{T} \int_0^L du {\hat{J}}'[b(u)] ( \tilde q_k - e^{iku}q_k )  
     - \frac{2}{T} \int_0^L du
     (1-e^{iku})\hat{J}'[\tilde{b}(u)]\tilde{q}_k \nonumber \\
     + 2 \int_{0}^{1} d\rho \int_0^L du e^{iku}{\hat{J}}'[\hat B^u(\rho)] \hat R_k(\rho)  
     + 4 \int_{0}^{1} d\rho \int_0^L du {\hat{J}}''[\hat B^u(\rho)] \hat R^u(\rho)  
          \left[ \tilde q_k - e^{iku}q_k C_k(\rho) \right] . 
  \label{eq:qtk1}  
\end{eqnarray} 
Eq.~(\ref{eq:CkTTI}) for $t\rightarrow\infty$ and then 
$\tau\rightarrow\infty$ results in 
\begin{eqnarray} 
  (\mu+Mk^2) q_k = \frac{2}{T} \int_0^L du (1-e^{iku})
  \{\hat{J}'[b(u)]-\hat{J}'[\tilde{b}(u)]\}q_k \nonumber \\
    + \frac{2}{T} \int_0^L du e^{iku}{\hat{J}}'[b(u)](\tilde q_k - q_k )  
     + 2 \int_{0}^{1} d\rho \int_0^L du e^{iku}{\hat{J}}'(\hat
     B^u(\rho)) \hat R_k(\rho)
     \nonumber \\  
     + 4 \int_{0}^{1} d\rho \int_0^L du {\hat{J}}''(\hat B^u(\rho)) \hat R^u(\rho)  
          q_k \left[ 1 - e^{iku}C_k(\rho) \right] . 
  \label{eq:qk1}  
\end{eqnarray} 
Also, Eq.(\ref{eq:CkAG}) for $\alpha\rightarrow 0$ gives 
\begin{eqnarray} 
  (\mu+Mk^2) q_{0,k} =  
     2 \int_0^L du e^{iku}{\hat{J}}'[b_0(u)] \int_{0}^{1} d\rho \hat R_k(\rho) +  
     \frac{2}{T} \int_0^L du e^{iku}{\hat{J}}'[b_0(u)] ( \tilde
     q_k - q_k )\nonumber \\
     \frac{2}{T}\int_0^L du (1-e^{iku})
  \{\hat{J}'[b(u)]-\hat{J}'[\tilde{b}(u)]\}q_{0,k} +
  \frac{4}{T}\int_0^L du (1-e^{iku})\hat{J}''[\hat{B}^u(\rho)]\hat{R}^u(\rho)q_{0,k}.
  \label{eq:q0k1}   
\end{eqnarray} 
Eqs.~(\ref{eq:qtk1}), (\ref{eq:qk1}) and (\ref{eq:q0k1}) and their origin 
Eqs.~(\ref{eq:CkTTI}), (\ref{eq:CkAG}) and (\ref{eq:RkAG}) contain both 
TTI and aging parts.  Thus, in principle, there are two ansätze for 
solving them, leading to two physical behaviors an ergodic phase (this
name was coined by KHF \cite{KHF} in the sense of non-glassy) characterized by TTI and FDT where
the aging behavior is completely missing, and a glassy phase
containing both stationary and aging behaviors.

\section{Ergodic Phase}

By An ergodic phase we mean that the external parameters are such that
only the stationary solution exists and not the aging solution. This
happens when the only solution is with $q_k=q_{0,k}$.
Technically, assuming that aging is absent amounts to setting $\hat 
R_k(\alpha)=0$ and $\hat C_k(\alpha)=1$ in (\ref{eq:qtk1}), 
(\ref{eq:qk1}) and (\ref{eq:q0k1}). (Equivalently, one could start 
from (\ref{eq:emC}) and (\ref{eq:emR}) and exclude the aging part from 
the beginning, leading to the same equations.)  Thus, in the ergodic 
phase, equations (\ref{eq:qtk1}), (\ref{eq:qk1}) and (\ref{eq:q0k1}) 
reduce to 
\begin{eqnarray} 
   && (\mu+Mk^2) \tilde q_k =  
      T + \frac{2}{T} \int_0^L du{\hat{J}}'[b(u)] (\tilde q_k -
      e^{iku}q_k)- \frac{2}{T}\int_0^L du (1-e^{iku})\hat{J}'[\tilde{b}(u)]\tilde{q}_k,  
   \label{eq:qter} \\ 
   && (\mu+Mk^2) q_k =  
      \frac{2}{T} \int_0^L du e^{iku}{\hat{J}}'[b(u)] (\tilde q_k
      - q_k) + \frac{2}{T} \int_0^L du (1-e^{iku})
  \{\hat{J}'[b(u)]-\hat{J}'[\tilde{b}(u)]\}q_k , 
   \label{eq:qker} \\ 
   && (\mu+Mk^2) q_{0,k} =  
      \frac{2}{T} \int_0^L du e^{iku}{\hat{J}}'[b_0(u)] (\tilde
      q_k - q_k) + \frac{2}{T} \int_0^L du (1-e^{iku})
  \{\hat{J}'[b(u)]-\hat{J}'[\tilde{b}(u)]\}q_{0,k} . 
   \label{eq:q0ker} 
\end{eqnarray} 
Note that (\ref{eq:qker}) and (\ref{eq:q0ker}) enforce $q_k=q_{0,k}$ 
which is just equivalent to $\hat C_k(\alpha)=1$, so one gets only 
two equations. 

In order to solve these equation we define the following constants,
which are themselves functions of $\tilde{q}_k$ and $q_k$:
\begin{eqnarray}
 E_k=\frac{2}{T} \int_0^L du e^{iku} \hat{J}'[b(u)],\\
 F_k= \frac{2}{T} \int_0^L du (1-e^{iku})\hat{J}'[\tilde{b}(u)].
\end{eqnarray}
In terms of these constants the equations for $\tilde{q}_k$ and $q_k$
become:
\begin{eqnarray}
  && (\mu+Mk^2) \tilde q_k =  
      T + E_0 \tilde q_k -E_k q_k - F_k \tilde{q}_k, \\
 && (\mu+Mk^2) q_k =  
      E_k (\tilde q_k
      - q_k) + (E_0-E_k-F_k) q_k . 
\end{eqnarray}
The solution of these equations is
\begin{eqnarray}
 \tilde{q}_k = \frac{T ( \mu + Mk^2-E_0+2E_k+F_k)}{ (\mu + Mk^2-E_0+
   E_k + F_k )^2}, \\
  {q}_k = \frac{T E_k}{ (\mu + Mk^2-E_0+E_k + F_k )^2}.
\end{eqnarray}
We now show that the ansatz $E_k=0$ for $k\neq0$ solves the
equations. Since $F_0=0$ we find
\begin{eqnarray}
  \tilde{q}_0=\frac{T}{\mu}+\frac{TE_0}{\mu^2},\\
  \tilde{q}_{k\neq 0}=\frac{T}{ \mu + Mk^2-E_0+F_k }\\
    q_0=\frac{TE_0}{\mu^2},\ \ \ q_{k\neq 0}=0.
\end{eqnarray}
Using these solutions we see that 
\begin{eqnarray}
  b(u)=\frac{2T}{L\mu}+\frac{2T}{L}\sum_{k\neq 0}\frac{1}{ \mu + Mk^2-E_0+F_k}\equiv b
\end{eqnarray}
is independent of $u$. Thus 
\begin{eqnarray}
 E_k= \frac{2}{T}\hat{J}'(b) \int_0^L du e^{iku} =
 \frac{2L}{T}\hat{J}'(b) \delta_{k,0}.
\end{eqnarray}
which validates our ansatz. Thus we can write
\begin{eqnarray}
 \label{eq:E0sol}
  E_0=\frac{2L}{T}\hat{J}'\left(\frac{2T}{L\mu}+\frac{2T}{L}\sum_{k\neq 0}\frac{1}{ \mu + Mk^2-E_0+F_k}\right)
\end{eqnarray}
Also
\begin{eqnarray}
  \label{eq:Fksol}
  \tilde{b}(u) &=&  \frac{2}{L}\sum_k (1-e^{-iku})\tilde{q}_k=\frac{2T}{L}\sum_k \frac{1-e^{-iku}}{ \mu + Mk^2-E_0+F_k } , 
\end{eqnarray}
and hence
\begin{eqnarray}
 F_k=  \frac{2}{T} \int_0^L du
 (1-e^{iku})\hat{J}'\left(\frac{2T}{L}\sum_{k'\neq 0}
   \frac{1-e^{-ik'u}}{ \mu + Mk'^2-E_0+F_{k'} }\right)
\label{eq:Fkforerg}
\end{eqnarray}
This equation together with Eq.(\ref{eq:E0sol}) gives a complete set of
equations in the ergodic case. They are identical to Eq.(4.15) and
(4.16) in Ref.~[\onlinecite{gold2}] derived by the replica method for a
quantum particle in a random potential(the notation there is slightly
different but it is easy to identify the corresponding
variables). Thus in the ergodic case there is a complete agreement
between the dynamical calculation and the replica calculation.

From the results above we obtain
\begin{eqnarray}
  \tilde{q}&=& \lim_{t\rightarrow\infty}\langle\langle
  x(s,t)x(s,t)\rangle\rangle/d =\frac{1}{L}\sum_k \tilde{q}_k
 =\frac{T}{L\mu}+\frac{TE_0}{L\mu^2}+
\frac{T}{L}\sum_{k\neq 0}\frac{1}{ \mu + Mk^2-E_0+F_k},\nonumber \\
  q &=& \lim_{\tau\rightarrow\infty}\lim_{t\rightarrow\infty}\langle\langle
  x(s,t+\tau)x(s,t)\rangle\rangle/d =\frac{1}{L}\sum_k q_k=\frac{TE_0}{L\mu^2}.
\label{eq:qergodic}
\end{eqnarray}

\section{Spin glass phase}
 
In this phase $q_0 \neq q_{0,k}$ and there is a time regime where the
aging behavior takes place. This phase corresponds to the RSB phase of
the statics.
We introduce the functions
\begin{eqnarray}
  E_{0,k}=\frac{2}{T} \int_0^L du e^{iku} \hat{J}'[b_0(u)], 
\end{eqnarray}
Keeping the aging parts and using the GFDT, Eqs.~(\ref{eq:qtk1}), 
(\ref{eq:qk1}) and (\ref{eq:q0k1}) can be transformed into 
\begin{eqnarray}
\label{eq:qtk2}
   (\mu+Mk^2) \tilde q_k =  
      T + E_0 (1-x_c) \tilde q_k -E_k (1-x_c) q_k - F_k \tilde{q}_k
      + E_{0,0} x_c \tilde{q}_k-E_{0,k} x_c q_{0,k}, \\
\label{eq:qk2}
   (\mu+Mk^2) q_k =  
      E_k (\tilde q_k - q_k) + [E_0(1-x_c)-E_k(1-x_c)+E_{0,0}x_c-F_k] q_k
      -E_{0,k}x_cq_{0,k},\\
\label{eq:q0k2}
 (\mu+Mk^2) q_{0,k} =  
      E_{0,k} \tilde q_k - E_{0,k}(1-x_c)q_k + [E_0(1-x_c)-E_k(1-x_c)+E_{0,0}x_c-2E_{0,k}x_c-F_k] q_{0,k}.
\end{eqnarray} 
In this case, similar to the ergodic case we will use the ansatz
$E_k=0$ and $E_{0,k}=0$ for $k\neq 0$. We will see that this provides
again a consistent solution. Using this ansatz the solution of 
Eqs.~(\ref{eq:qtk2}), (\ref{eq:qk2}) and (\ref{eq:q0k2}) for 
$\tilde q_k$, $q_k$ and $q_{0,k}$ becomes
\begin{eqnarray}
   \tilde{q}_0=\frac{T}{\mu+\Sigma}+q_0,\ \ \ \tilde{q}_{k\neq
   0}=\frac{T}{ \mu + Mk^2-E_0+\Sigma+F_k } \\
    q_0=\frac{T(\mu E_0+E_{0,0}\Sigma)}{\mu^2(\mu+\Sigma)},\ \ \
    q_{k\neq 0}=0, \\
    q_{0,0}=\frac{T E_{0,0}}{\mu^2}\ \ \ \ q_{0,k\neq 0}=0. 
\label{eq:sgsol}
\end{eqnarray}%
where 
\begin{equation} 
   \Sigma = x_c (E_0-E_{0,0}). 
\end{equation} 
Using this solution we see that
\begin{eqnarray} 
   {b}(u) &=& \frac{2T}{L(\mu+\Sigma)}+\frac{2T}{L}\sum_{k\neq 0}\frac{1}{\mu+M
     k^2-E_0+\Sigma+F_k} \equiv b,\\ 
  {b_0}(u) &=& \frac{2T(\mu+E_0-E_{0,0})}{L\mu(\mu+\Sigma)}+\frac{2T}{L}\sum_{k\neq
   0}\frac{1}{\mu+M k^2-E_0+\Sigma+F_k} \equiv b_0,    
\end{eqnarray} 
both independent of $u$. Thus
\begin{eqnarray}
   E_k= \frac{2}{T}\hat{J}'(b) \int_0^L du e^{iku} =
 \frac{2L}{T}\hat{J}'(b) \delta_{k,0},\\
 E_{0,k}= \frac{2}{T}\hat{J}'(b_0) \int_0^L du e^{iku} =
 \frac{2L}{T}\hat{J}'(b_0) \delta_{k,0},
\end{eqnarray}
consistent with our ansatz. If we denote 
\begin{eqnarray}
  a(u)=\frac{2T}{L}\sum_{k\neq 0}\frac{e^{-iku}}{\mu+M k^2-E_0+\Sigma+F_k},
\label{eq:au}
\end{eqnarray}
we can write
\begin{eqnarray}
  E_0=\frac{2L}{T}\hat{J}'\left[a(0)+\frac{2T}{L(\mu+\Sigma)}\right],\label{eq:E0eq}\\
  E_{0,0}=\frac{2L}{T}\hat{J}'\left[a(0)+\frac{2T}{L(\mu+\Sigma)}\left(1+\frac{\Sigma}{\mu x_c}\right)\right].
\label{eq:E00eq}
\end{eqnarray}
and thus
\begin{eqnarray}
  \Sigma= \frac{2L x_c}{T}
  \left\{\hat{J}'\left[a(0)+\frac{2T}{L(\mu+\Sigma)}\right]
-\hat{J}'\left[a(0)+\frac{2T}{L(\mu+\Sigma)}\left(1+\frac{\Sigma}{\mu x_c}\right)\right]\right\}
\label{eq:Sigmaeq}
\end{eqnarray}
For $F_k$ we get instead of Eq.(\ref{eq:Fkforerg})
\begin{eqnarray}
 F_k=  \frac{2}{T} \int_0^L du
 (1-e^{iku})\hat{J}'\left[a(0)-a(u)\right]
\label{eq:Fkeq}
\end{eqnarray}
Let us define the functions 
\begin{eqnarray}
  D_k=\frac{2}{T} \int_0^L du e^{iku} \hat{J}''[b(u)]\hat{R}^u(1),
\end{eqnarray}
from Eq.(\ref{eq:GFDT}) and the fact that $q_{k\neq 0}=0$ it follows 
that $\hat{R}_{k\neq
0}(\alpha)=0$, and hence also $D_{k\neq 0}=0$. Thus 
\begin{eqnarray}
  D_k= \frac{2}{T} \hat{J}''[b]\hat{R}_0(1) \delta_{k,0}.
\end{eqnarray}
Furthermore, Eq.~(\ref{eq:RkAG}) with $\alpha=1$ and $k=0$ gives 
\begin{equation} 
   (\mu + \Sigma) \hat R_0(1) = - 2 D_0 (\tilde q_0 - q_0 )
=-\frac{4}{\mu+\Sigma}\hat{J}''(b)\hat R_0(1),
\end{equation} 
which can be written as 
\begin{equation} 
  0 = \hat R_0(1) \left[ 1 + \frac{4}{(\mu+\Sigma)^2} {\hat{J}}''(b) \right]. 
  \label{eq:hatr} 
\end{equation} 
Eq.~(\ref{eq:hatr}) with $\hat R_0(1)\ne 0$ implies the
marginal stability condition 
\begin{equation} 
  - \left(\frac{T}{L}\right)^2 =
  \left[\frac{2T}{L(\mu+\Sigma)}\right]^2 {\hat{J}}''
  \left[a(0)+\frac{2T}{L(\mu+\Sigma)}\right]. 
  \label{eq:margin}  
\end{equation}  
Eqs.~(\ref{eq:Sigmaeq}), (\ref{eq:lambdaeq}), (\ref{eq:aulambda})
and (\ref{eq:margin}) fully solve the model. These equations, with the
exception of Eq.(\ref{eq:margin}), are the same as the equations for
the statics that we obtained using the replica method in
Ref.~[\onlinecite{gold2}] and
Ref.~[\onlinecite{gold}]. Equation (\ref{eq:margin}) though, is of pure
dynamic origin, and differs from the corresponding equation obtained
in the replica method by extremizing the variational free energy with
respect to the RSB breakpoint $x_c$. We give here for comparison the corresponding
equation obtained in the statics replica calculation \cite{gold}:
\begin{eqnarray}
\label{eq:staticsx}
\frac{L^2 x_c^2}{T^2}
  \left\{\hat{J}\left[a(0)+\frac{2T}{L(\mu+\Sigma)}\right]
-\hat{J}\left[a(0)+\frac{2T}{L(\mu+\Sigma)}\left(1+\frac{\Sigma}{\mu
      x_c}\right)\right]\right\} \nonumber \\
+ \frac{2L x_c}{T}\frac{\Sigma}{\mu(\mu+\Sigma)}
  \hat{J}'\left[a(0)+\frac{2T}{L(\mu+\Sigma)}\left(1+\frac{\Sigma}{\mu
      x_c}\right)\right]-\frac{\Sigma}{\mu+\Sigma}+\ln\left(1+\frac{\Sigma}{\mu}\right)=0. 
\end{eqnarray}
Equation (\ref{eq:margin}) can  be obtained in the replica calculation
by requiring marginal stability i.e. the condition of a vanishing
replicon mass, but it does not correspond to the optimized variational
solution.

\section{Solving the equations} 
 
In this section we discuss the solution to the equations  derived in
the last section. Although a full solution can be found numerically,
our goal here is to go as far as one can with analytic methods because
it gives a better understanding of the nature of the solution. An
analytical solution becomes possible for a system of large volume
(small value of $\mu$) and whwn the polymer is very long (large $L$).

Before we consider the solution of the equations derived above for a
polymer, let us first discuss the special limit where the problem
reduces to a classical particle in a random potential. Looking back at
the original hamiltonian one realizes that in the limit
$M\rightarrow \infty$ and $T\rightarrow TL$ (or alternatively $L=1$) 
the problem should become the same as the particle problem discussed in
Refs.~[\onlinecite{engel,KH1,CD}]. Looking back at our
equations we see that in the limit $M \rightarrow \infty$ only the
$k=0$ component of $\tilde q_k$ survive. 

In the ergodic case $E_0$ is given by
\begin{eqnarray}
E_0=\frac{2}{T}\hat{J}'\left(\frac{2T}{\mu}\right),
\end{eqnarray}
and the expressions (\ref{eq:qergodic}) for $\tilde q$ and $q$ become
\begin{eqnarray}
  \tilde{q}
 &=&\frac{T}{\mu}+\frac{TE_0}{\mu^2},\\
  q &=& \frac{TE_0}{\mu^2}.
\end{eqnarray}
These are exactly the same as equations (57)-(59) derived by Engel
\cite{engel} using the replica method and correspond to his replica
symmetric solution.

In the spin glass case we again see that as $M \rightarrow \infty$ we have
$a(u)=0$. Thus Eqs.(\ref{eq:E0eq}),(\ref{eq:E00eq})
and (\ref{eq:Sigmaeq}) with $a(0)=0$ and $L=1$ agree exactly with equations
(74)-(76) of Engel \cite{engel} using replicas with
1-step RSB . However it is Eq.(\ref{eq:staticsx}) with $a(0)=0$ and $L=1$ that
agrees with Engel's Eq.(78) and not the dynamical equation
(\ref{eq:margin}) with $a(0)=0$ that agrees with the dynamical
equation derived in Refs.~[\onlinecite{CD,KH1}]. We also
find that 
\begin{eqnarray}
 \tilde q = \frac{T}{\mu+\Sigma}\left(1+\frac{E_0}{\mu}+\frac{E_{0,0}\Sigma}{\mu^2}\right) 
\end{eqnarray}
agree with Engel's Eq.(79).

Let us review Engel's solution to the variational equations for the
short range, random potential case in the
limit of small $\mu$ (but still $\mu \neq 0$). Eq.(\ref{eq:staticsx})
with $a(0)=0$ and $L=1$ gives
\begin{eqnarray}
 \frac{x_c^2}{T^2}\hat{J}\left(\frac{2T}{\Sigma}\right)-\ln(\mu)=0, 
\label{eq:xcforcp}
\end{eqnarray}
Thus
\begin{eqnarray}
  x_c=T \left(\frac{|\ln(\mu)|}{|\hat{J}(0)|}\right)^{1/2},
\end{eqnarray}
 where we used the fact that $2T/\Sigma\rightarrow 0$ for
$\mu\rightarrow 0$ as will become clear shortly. Notice that in this
case of a particle in a random potential we must consider a random
potential that is regularized at small distances (not a strict
$\delta$-function), so $\hat J (0)$ is properly defined. Equation
(\ref{eq:Sigmaeq}) then gives
\begin{eqnarray}
  \Sigma = \frac{2 x_c}{T} \hat{J}'(0)= 2\hat{J}'(0)\left(\frac{|\ln(\mu)|}{|\hat{J}(0)|}\right)^{1/2}
\end{eqnarray}
Also $E_{0,0}$ vanishes like 
\begin{eqnarray}
  E_{0,0}=\frac{2}{T}\hat J'\left(\frac{2T}{\mu x_c}\right)\sim
  \frac{g}{T(2\pi)^{d/2}}\left(\frac{\mu^2 |\ln(\mu)|}{2|\hat J(0)|}\right)^{(2+d)/4}
\end{eqnarray}
and $E_0=2\hat J'(0)/T$. Thus  
\begin{eqnarray}
 \tilde q \approx \frac {T E_0}{\Sigma \mu}\sim \frac{|\hat
   J(0)|^{1/2}}{\mu \sqrt{|\ln(\mu)|}}
\label{eq:IM}
\end{eqnarray}
The importance of Eq.(\ref{eq:IM}) is that it is consistent with the
following Imry-Ma type argument:
Since the random potential has a gaussian distribution of variance
$g$, in a volume of radius $r_g$ the
minimum of the potential is given by $V_m\sim -\sqrt{g\ln(r_g)}$. This result
follows from the fact that when we pick $r_g^d$ numbers from a gaussian
distribution with zero mean and unit variance, the lowest expected
value of the potential is given by the solution of the equation
\begin{eqnarray}
  \int_{-\infty}^{V_m} \exp{(-V^2/2g)} \sim \frac{1}{r_g^d}.
\end{eqnarray}
 Thus the Hamiltonian is $H(x_g)\sim (1/2)\mu r_g^2-\sqrt{g\ln{r_g}}$
 and minimization gives 
 \begin{eqnarray}
  r_g \sim \frac{g^{1/4}}{\sqrt{\mu}|\ln{\mu}|^{1/4}}, 
 \end{eqnarray}
yielding
\begin{eqnarray}
  \tilde q = \langle\langle x^2\rangle\rangle \sim \frac{g^{1/2}}{\mu\sqrt{|\ln{\mu}|}}.
\end{eqnarray}
This is a nonanalytic expression that cannot be obtained from
perturbation theory. The critical temperature below which the RSB
solution is the stable one is given by the condition $x_c=1$ or $T_c =
(\hat J(0)/|\ln \mu |)^{1/2}$.

In the dynamics approach though, Eq.~(\ref{eq:margin})
replaces Eq.~(\ref{eq:staticsx}). For small $\mu$ since $\Sigma \gg \mu$
Eq.~(\ref{eq:margin}) gives
\begin{eqnarray}
  -1=\frac{4}{\Sigma^2}\hat J''(0),
\end{eqnarray}
or
\begin{eqnarray}
  \Sigma=2|\hat J''(0)|^{1/2}.
\end{eqnarray}
again $E_0=2\hat J'(0)/T$ and $E_{0,0}$ vanishes like $\mu^{(2+d)/2}$
and thus from Eq.~(\ref{eq:Sigmaeq}) we get 
\begin{eqnarray}
  x_c= T\frac{|\hat J''(0)|^{1/2}}{\hat J'(0)}.
\end{eqnarray}
We also find that 
\begin{eqnarray}
 \tilde q \approx \frac {T E_0}{\Sigma \mu}\sim \frac{\hat
   J'(0)}{\mu |\hat J''(0)|^{1/2}}.
\label{eq:dynmsd}
\end{eqnarray}
We observe that in the dynamical formulation $\ln \mu $ is replaced by
the constant $\hat J''(0) |\hat J(0)|/(\hat J'(0))^2$ which is equal
to $(d+2)/d$ for the short ranged correlated potential. The Imry-Ma result
is not satisfied, even though the distance of the particle from the
origin still diverges for $\mu\rightarrow 0$. This is probably due
to the fact that in the
dynamical formulation the particle is impeded by large barriers to
search the entire volume as effectively for the lowest minimum of the
random potential as is obtained in the statics, even in the limit of
large times. The fact that the dynamical solution differs from the
statics in this case was observed in Ref.~[\onlinecite{CD}], but they did not compare
the solutions in detail. They observe that in the dynamical solution
the free energy is higher than the replica free energy obtained in the
statics. This discrepancy may also be due to the fact that the initial
conditions of the dynamics are completely random and are not weighted
with an appropriate Boltzmann factor \cite{houghton}. 

We now return to the polymer problem and discuss the solution
to the previously derived equations for the case of a
random potential with short range correlations characterized
by $\hat \gamma=1+d/2$. The ergodic phase for large times in principle
should correspond to the replica symmetric solution of the statics. In
the equilibrium solution such a solution arises in two situations: First,
in the infinite volume limit, when $\mu =0$. In this case the polymer
totally collapses to a size of a single monomer, since deep minima of 
unbounded bottom exist in this limit. This case was discussed 
before in Ref.~[\onlinecite{gold}] and it was shown that the 
solution corresponds to the case of an annealed disorder.
Since the solution was discussed before we will not repeat it here. 
In this case the equations for the
dynamics are the same as obtained in the replica solution. 
Second, for the finite volume
case the ergodic solution is applicable when $L<L_c$ where $L_c$ is
the value for which the spin glass solution ceases to exist, see
below. It turns out that $L_c$ is a number of order unity, thus except
for very short chains the non-ergodic solution is the correct one.
It is a subtle question if the domain of validity of the ergodic phase
in the dynamics and in the statics coincide. For the case of infinite
volume one has to take the limit $\mu\rightarrow 0$ before the limit of
large time. we will not discuss this question further here.

In the Appendix we give the
solution of the ergodic equations for a special case of a long range random potential
with quadratic correlations and show that it gives the same solution
as previously obtained by Shiferaw and Goldschmidt \cite{shifgold2} using a
different method. Of course for a quadratically correlated potential
the variational approximation is exact.

We proceed to discuss the solution to the equations  in the spin-glass
case when the random potential has short range correlations ($\hat\gamma=1+d/2$). As mentioned above, 
when the volume is large but finite the spin-glass
solution is the appropriate solution when $L$ is large enough \cite{gold}. For the spin glass case
we have seen that one of the equations of the dynamics differs from one of the equations
derived in the statics using the replica method. We now re-derive
the solution of the equations of the statics and then show how the
solution of the equations of the dynamics differs from it. The reason
that we reconsider the solution of the equations of the statics is
first because the variational scheme that emerged above is more
general than the scheme employed in Ref.~[\onlinecite{gold}] since it
involves more variational parameters like the scheme used in
Ref.~[\onlinecite{gold2}], and we want to show that the end results are
the same. Second, the steps of the solution will facilitate the
solutions of the dynamical equations.
 
In the limit of small $\mu$ (but still $\mu \neq 0$). Eq.(\ref{eq:staticsx})
gives
\begin{eqnarray}
 \frac{x_c^2 L^2}{T^2}\hat{J}\left(a(0)+\frac{2T}{L\Sigma}\right)-\ln(\mu)=0, 
\label{eq:xcforpol}
\end{eqnarray}
Thus
\begin{eqnarray}
  x_c=\frac{T}{L} \left(\frac{|\ln(\mu)|}{|\hat{J}[a(0)]|}\right)^{1/2},
\label{eq:xcsg}
\end{eqnarray}
 where we used the fact that $2T/(L\Sigma) \ll a(0)$ for large $L$ ( and
 even for finite $L$ as $\mu\rightarrow 0$ as can be checked
 a posteriori with full solution). Equation
(\ref{eq:Sigmaeq}) then gives
\begin{eqnarray}
  \Sigma = \frac{2 L x_c}{T} \hat{J}'[a(0)]= 2\hat{J}'[a(0)]\left(\frac{|\ln(\mu)|}{|\hat{J}[a(0)]|}\right)^{1/2}
\label{eq:Sigmasg}
\end{eqnarray}
and from Eq.(\ref{eq:E00eq}) one obtains
\begin{eqnarray}
  E_{0,0}=\frac{2L}{T}\hat J'\left(\frac{2T}{\mu L x_c}\right)\sim
  \frac{gL}{T(2\pi)^{d/2}}\left(\frac{\mu^2 |\ln(\mu)|}{4|\hat J[a(0)]|}\right)^{(2+d)/4}.
\end{eqnarray}
It is convenient to define the variables
\begin{eqnarray}
\label{eq:lambda}
 \lambda_k = F_k - E_0 +\Sigma +\mu,\ \ \ \ \ \ k \neq 0, 
\end{eqnarray}
Using Eq.(\ref{eq:E0eq}), $\lambda_k$ satisfy the equation
\begin{eqnarray}
\label{eq:lambdaeq}
  \lambda_k =  \mu + \Sigma +\frac{2}{T} \int_0^L du
 \left\{\left(1-e^{iku}\right)\hat{J}'\left[a(0)-a(u)\right]\right.\nonumber \\
 \left.- \hat{J}'\left[a(0)+\frac{2T}{L(\mu+\Sigma)}\right]\right\}.
\end{eqnarray}
and in terms of $\lambda_k$ the constant $a(u)$ is given by
\begin{eqnarray}
\label{eq:aulambda}
 a(u)= \frac{2T}{L}\sum_{k\neq 0}
   \frac{e^{-iku}}{ Mk^2 + \lambda_{k} }\rightarrow
   2T\int_{-\infty}^\infty \frac{dk}{2\pi}
   \frac{e^{-iku}}{ Mk^2 + \lambda_{k} }
\end{eqnarray}
where the last expression is valid for large $L$. 
The integral on the right hand side of Eq.~(\ref{eq:lambdaeq}) converges
as $L$ becomes large and thus to leading order the parameters $\lambda_k$ are O(1) with
respect to $L$. To simplify the integral further notice first that the
integrand is invariant under the transformation $u \rightarrow L-u$
thus it symmetric about $u=L/2$. The total integral is twice its value up 
to $u=L/2$ and $\lambda_k$ can consistently be taken as real. For large
$L$ we obtain 
\begin{eqnarray}
\label{eq:lambdakL}
  \lambda_k = \mu +\Sigma-\frac{4}{\Sigma}\hat J''[a(0)]+ \frac{4}{T} \int_0^\infty du
 \left(1-\cos{ku}\right)\left\{\hat{J}'\left[a(0)-a(u)\right]
 - \hat{J}'\left[a(0)\right]\right\}+O(1/L),
\end{eqnarray}
where $\Sigma$ is given by Eq.(\ref{eq:Sigmasg}) and 
\begin{eqnarray}
  a(u) = 2T\int_{-\infty}^\infty \frac{dk}{2\pi}
   \frac{\cos{ku}}{ Mk^2 + \lambda_{k} }
\label{eq:ausym}
\end{eqnarray}
Note that we have added to the integral in Eq.(\ref{eq:lambdakL}) a
term
\begin{eqnarray}
  \frac{4}{T}\hat J'[a(0)]\int_0^L du \cos(ku) =0 , \ \ \ \ k\neq 0.
\end{eqnarray}
Our goal is to characterize the behavior of the end-to-end distance of
the polymer that can be extracted from the correlation function
\begin{eqnarray}
  \label{eq:cf}
  \tilde b(L)= \lim_{t\rightarrow \infty} \langle\langle
  [x(L,t)-x(0,t)]^2\rangle\rangle = \frac{2}{L} \sum_{k\neq0}(1- e^{-ikL})\tilde q_k=a(0)-a(L).
\end{eqnarray}
We are now going to argue that as $\mu\rightarrow 0$ (but still
finite), $\lambda_k$ satisfies, to leading order in $|\ln\mu|$ the scaling form 
\begin{eqnarray}
  \label{eq:lambdakasym}
  \lambda_k= (g|\ln\mu|)^\delta \tilde \lambda +
  \frac{(g|\ln\mu|)^\delta}{|\ln\mu|} f\left(k
    (g|\ln\mu|)^{-\delta/2}\right),  \ \ \ k>0
\end{eqnarray}
where $\delta=4/(4-d)$ and $\lambda_{-k}=\lambda_k$. For small $k$ the function $f$
satisfies $f(x)\sim x^2$ and is regular around 0. For large x, it can be
shown that $f(x)\sim
x^{d/2}$. Substituting Eq.(\ref{eq:lambdakasym}) into Eq.(\ref{eq:ausym}) and changing the
integration variable  
\begin{eqnarray}
 k\rightarrow k (g|\ln \mu|)^{\delta/2},
\end{eqnarray}
we find
\begin{eqnarray}
   a(u) = \frac{2T}{(g|\ln \mu|)^{\delta/2}}\int_{-\infty}^\infty \frac{dk}{2\pi}
   \frac{\cos(k(g|\ln \mu|)^{\delta/2}u)}{ Mk^2 + \tilde \lambda+\frac{1}{|\ln\mu|}f(k) }
\end{eqnarray}
From which it follows that to leading order in $|\ln\mu|$
\begin{eqnarray}
  a(u)=\frac{T}{\sqrt{\tilde\lambda
      M}}(g|\ln \mu|)^{-\delta/2}e^{-\sqrt{\tilde\lambda/M}(g|\ln
    \mu|)^{\delta/2}u}
\label{eq:aulnmu}
\end{eqnarray}
and in particular $a(0)=T/\sqrt{M\tilde\lambda}\ (g|\ln
\mu|)^{-\delta/2}$. Substituting on the rhs of Eq.(\ref{eq:lambdakL}) and
changing the integration variable
\begin{eqnarray}
 u\rightarrow u \ (g|\ln \mu|)^{-\delta/2}(M/\tilde\lambda)^{1/2},
\end{eqnarray}
we find for small $\mu$
\begin{eqnarray}
  \lambda_k=\frac{d^{1/2}}{(2\pi)^{d/4}}\left(\frac{T}
{\sqrt{M\tilde\lambda}}\right)^{-(d+4)/4}(g|\ln\mu|)^{1/2+
\delta(d+4)/8}\left(1+O\left(\frac{1}{|\ln\mu|}\right)\right)+\nonumber\\
\frac{(g|\ln\mu|)^{1+\delta d/4}}{|\ln\mu|}f\left(k (g|\ln\mu|)^{-\delta/2}\right),
\end{eqnarray}
where
\begin{eqnarray}
 f(x)=\frac{2M}{(2\pi)^{d/2}T^{2}}\left(\frac{T}{\sqrt{M\tilde\lambda}}\right)^{-d/2}
\int_0^\infty du
\left\{1-\cos\left[x(M/\tilde\lambda)^{1/2}u\right ]\right\} 
\left\{\frac{1}{\left(1-e^{-u}\right)^{d/2+1}}-1\right\}. 
\end{eqnarray}
we see that for consistency $\delta$ must satisfy $\delta=4/(4-d)$. 
We also see that the parameter $\tilde \lambda$ is given by
\begin{eqnarray}
 \tilde\lambda=\left(\frac{d}{(2\pi)^{d/2}}\right)^{4/(4-d)}\left(\frac{M}{T^2}\right)^{(4+d)/(4-d)}. 
\end{eqnarray}
Using Eq.(\ref{eq:aulnmu}) we have
\begin{eqnarray}
  \tilde b(L)=\frac{T}{\sqrt{\tilde\lambda
      M}}(g|\ln \mu|)^{-\delta/2}\left(1-e^{-\sqrt{\tilde\lambda/M}(g|\ln
    \mu|)^{\delta/2}L}\right).
\label{eq:CFL}
\end{eqnarray}
Thus the end-to-end distance is given by
\begin{eqnarray}
  R^2 \sim \frac{T}{\sqrt{\tilde\lambda
      M}}(g|\ln \mu|)^{-\delta/2},
\end{eqnarray}
from which it follows that 
\begin{eqnarray}
  R \sim (g|\ln \mu|)^{-1/(4-d)}\sim (g\ln {\cal V})^{-1/(4-d)}.
\end{eqnarray}
Here $\cal V$ is the total volume, and this result has the correct $g$
dependence and the subtle $\ln \cal V$ dependence as was originally
argued by Cates and Ball \cite{cates} and derived in
Ref.~[\onlinecite{gold}] using the replica method.

The important observation is that when $k$ increases from 0 to
$\sqrt{\tilde \lambda/M}(g|\ln \mu)^{\delta/2}$, $\lambda_k$ changes only by a
factor $1+O(1/|\ln\mu|)$ and thus the results of the single $\lambda$
parameter used in the variational scheme of Ref.~[\onlinecite{gold}]
remain intact.  The effective mass of the low lying non-zero modes is
approximately given by $\sqrt{\lambda_k}\sim \sqrt{\Sigma}$.
   
 The above scaling arguments are valid
provided $a(0)>d\sigma^2/2$ (see Eq.(\ref{Jhatgauss}) which is what we are
going to assume. Of course for a delta function correlated random
potential $\sigma\rightarrow 0$ and this condition always holds.
For this condition to apply in the case of other short range
correlated random potentials, $g$ the variance of the disorder, may not
be too large, such that the size of the polymer is not smaller than
the correlation length of the disorder.

The parameters $\Sigma$ and $x_c$ are given to leading order in $L$
and $|\ln\mu|$ by
\begin{eqnarray}
  \Sigma= d^{4/(4-d)}(2\pi)^{-2d/(4-d)}(M/T^2)^{(4+d)/(4-d)} (g |\ln \mu|)^{4/(4-d)},
\end{eqnarray}

\begin{eqnarray}
  x_{c}=\frac{1}{L}\left(  \frac{d^{d-2}}{(2\pi)^{d}}g^{2}T^{-(d+4)}%
M^{d}\left|  \ln\mu\right|  ^{d-2}\right)  ^{-1/(4-d)}.
\end{eqnarray}
We see that $x_c<1$ for large enough $L$, and for $d>2$ for fixed $L$
and small enough $\mu$. The value of $L$ corresponding to $x_c=1$ is
denoted by $L_c$. The condition $x_c<1$ can be written as
\begin{eqnarray}
  \label{eq:localization}
  \ln {\cal V} < L \sqrt{\frac{g \ln {\cal V}}{R^d}},
\end{eqnarray}
which means that the translational entropy is smaller than the binding
energy in the typical minimum of the random potential, thus implying
that the polymer is truly localized. It also follows that $x_c<1$
is equivalent to the condition
\begin{eqnarray}
  \frac{gL^{(4-d)/2}}{T^2b_K^d}(\ln {\cal V})^{(d-2)/2}>1,
\end{eqnarray}
up to some unimportant numerical factor.

We now discuss the solution of the dynamical equations. Eq.~(\ref{eq:margin})
replaces Eq.~(\ref{eq:staticsx}). For small $\mu$ since $\Sigma \gg \mu$
Eq.~(\ref{eq:margin}) gives
\begin{eqnarray}
  -1=\frac{4}{\Sigma^2}\hat J''[a(0)],
\end{eqnarray}
or
\begin{eqnarray}
  \Sigma=2\left|\hat J''[a(0)]\right|^{1/2}.
\end{eqnarray}
From Eq.~(\ref{eq:Sigmaeq}) we get 
\begin{eqnarray}
  x_c= \frac{T|\hat J''[a(0)]|^{1/2}}{L\hat J'[a(0)]},
\end{eqnarray}
and we see that there is no longer any $\ln \mu$ dependence as in the
replica solution. Again $x_c<1$ for large enough $L$. 
Using this equation we see that  
\begin{eqnarray}
  E_{0,0}=\frac{2L}{T}\hat J'\left(\frac{2T}{\mu L x_c}\right)\sim
  \frac{gL}{T(2\pi)^{d/2}}\left(\frac{\mu^2 |\hat J''[a(0)]|}{4\hat J'[a(0)]^2}\right)^{(2+d)/4}.
\label{eq:dynE00}
\end{eqnarray}
The equation for $\lambda_k$ again reads
\begin{eqnarray}
\label{eq:lambdamu0dyn}
  \lambda_k = \mu +\Sigma-\frac{4}{\Sigma}\hat J''[a(0)]+ \frac{4}{T} \int_0^\infty du
 \left(1-\cos{ku}\right)\left\{\hat{J}'\left[a(0)-a(u)\right]
 - \hat{J}'\left[a(0)\right]\right\}+O(1/L),
\end{eqnarray}
Since in this case there in no $\ln \mu$ dependence we can repeat the
scaling of $\lambda_k$ with $g$ as before but not with $\ln \mu$. We
find the $g$ dependence as before, thus
\begin{eqnarray}
  \lambda_k=g^{4/(4-d)}\tilde \lambda_k,
\end{eqnarray}
 where $\tilde\lambda_k$ are independent of $g$. We also find that
 since $|\ln \mu|$ is absent, the parameter $M$ will be
shifted by contributions from $f(k)$. We can still argue that
 $\lambda_k$ are independent of $L$ which is an important result, and
 that they have the $g$ dependence described above. This implies that 
 \begin{eqnarray}
   R\sim g^{-1/(4-d)}
 \end{eqnarray}
independent of $L$. Similarly we find that $\Sigma$ is finite and satisfies
 $\Sigma \sim g^{4/(4-d)}$ and $x_c\sim (1/L)g^{-2/(4-d)}$. This
last condition implies that the polymer is localized whenever $x_c<1$
which means approximately  $gL^{(4-d)/2}>1$.

These results still contain important physics, namely that the size of the
polymer is independent of its length and also it has the correct
dependence on the strength of the disorder, but the more subtle $\ln
\cal V$ 
dependence resulting from a sophisticated Imry-Ma type
argument is missing. 

\section{Discussion of the dynamical behavior}

Here we make some further comments about the dynamical behavior of the
polymer in the spin glass phase. Consider the motion of the center of 
mass of the polymer:
\begin{eqnarray}
B_{CM}(t,t')\equiv\langle\langle [x_{CM}(t)-x_{CM}(t')]^2\rangle\rangle/d=
\nonumber \\ \frac{1}{dL^2}\int_0^L du \int_0^L du'
\langle\langle[x(u,t)-x(u,t')]\cdot[x(u',t)-x(u',t')]\rangle\rangle=
\nonumber \\
\frac{1}{L}\left[C_0(t,t)+C_0(t',t')-2C_0(t,t')\right].
\end{eqnarray}
Let us denote $t'=t_W$ and $\tau=t-t'$. In the TTI regime for
large $t_W$ and $\tau \ll t_W$, as $\tau$ increases, $B_{CM}(\tau)$
increases from $0$ to
\begin{eqnarray}
b^{(1)}=\lim_{\tau\rightarrow\infty} \lim_{t_W\rightarrow
\infty}B_{CM}(t_W+\tau,t_W)=\frac{2}{L}\left(\tilde q_0-q_0\right)=\frac{2T}{L(\mu+\Sigma)}.
\end{eqnarray}
In this regime the FDT also holds. Let us compare this behavior with
the behavior of a free chain. For a free chain, for large $t,t'$ we have
\begin{eqnarray}
  B_{CM}(\tau)= \frac{2T}{L\mu}\left(1-e^{-\mu\tau}\right)\rightarrow_{\tau\rightarrow\infty} \frac{2T}{L\mu},
\end{eqnarray}
and thus when disorder is present $\Sigma$ is replacing $\mu$ when
$\mu\rightarrow 0$ and we observe that $b^(1)$ remains finite
as $\mu\rightarrow 0$. In this regime the polymer becomes trapped inside a
local potential minimum and we see that for large $\tau$ ( but still
less than $t_W$) there is no diffusion, even though for very small
$\tau$ we expect diffusive behavior. Thus for large enough $\tau$ 
there will be a plateau in the plot of $B_{CM}(\tau)$ as a function of
$\tau$. In this regime the short time estimates of previous
investigations (see e.g. Ref.~[\onlinecite{ebert2})] should be valid. 

However, for large and fixed $t_W$ as $\tau$ becomes sufficiently large,
$B_{CM}$ will leave the plateau and continue to grow above $b^{(1)}$
until it reaches the value
\begin{eqnarray} 
b^{(0)}=\lim_{\tau\rightarrow\infty}B_{CM}(t_W+\tau,t_W)=\frac{2}{L}(\tilde
q_0-q_{0,0})=\frac{2T}{L(\mu+\Sigma)}\left(1+\frac{\Sigma}{\mu x_c}\right).
\end{eqnarray} 
The size of the plateau depends on $t_w$. The larger $t_W$ the larger the
plateau, and the larger value of $\tau$ required for $B_{CM}$ to
increase beyond $b^{(1)}$. Thus the polymer does not remain trapped
forever but eventually hops to another minimum of the potential. Notice also that
 as $\mu\rightarrow 0$, $b^{(0)}\rightarrow 2T/(\mu L x_c)\sim 2Tg^{2/(4-d)}/\mu$. This value
 is independent of $L$ and should represent the typical square of the
 hopping distance of the polymer among different local minima of the
 potential. The larger the waiting time the deeper the local minimum
 occupied by the polymer and hence the longer it takes it to hop to
 another minimum. 
From with this observation we still lack an estimate of the time
dependence of $B_{CM}(t,t')$ in the vicinity of $b^{(1)}$ and 
$b^{(0)}$. A more detailed calculation is needed to derive the asymptotic 
growth rate in the different time regimes, and this will be left for future work.

Another quantity of interest is the single segment (bead or monomer)
mean square displacement
\begin{eqnarray}
B^{u=0}(t,t')=\langle\langle[x(s,t)-x(s,t')]^2\rangle\rangle/d=\nonumber\\
\frac{1}{L} \sum_k \left[C_k(t,t)+C_k(t',t')-2C_k(t,t')\right]. 
\end{eqnarray}
In this case the asymptotic mean square displacement in the TTI
regime becomes
\begin{eqnarray}
  b_{seg}^{(1)}=2(\tilde q-q)=b^{(1)}+a(0),
\end{eqnarray}
and the asymptotic value in the aging regime becomes
\begin{eqnarray}
  b_{seg}^{(0)}=2(\tilde q-q_0)=b^{(0)}+a(0),
\end{eqnarray}
with $b^{(1)}$ and $b^{(0)}$ defined above and
\begin{eqnarray}
  a(0)\sim (Tb_K^2)^{(4/(4-d)} g^{-2/(4-d)}.
\end{eqnarray}
The quantity $b_{seg}^{(1)}$ is dominated by $a(0)$ and thus the size
of the mean square displacement of a bead is the same as the square of
the end-to-end distance. The quantity $b_{seg}^{(0)}$ remains
essentially the same as $b^{(0)}$ for small $\mu$. 

The results of this section i.e. the width of a localized state of the
polymer and the average distance squared between different localized
states is the same as discussed in the interpretation of the 1-step RSB
solution in Ref.[~\onlinecite{shifgold2}] section VII, with only the 
subtle dependence on $\ln {\cal V}$ missing. 

\section{Conclusions} 
 In this paper we derived the dynamical equation for a gaussian chain
 in a short range correlated random potential. We used a simple
 Langevin dynamics and we discovered that there are two possible
 scenarios at large times: an stationary regime where the FDT applies and
 an aging regime where the FDT breaks down at large time separation. 
In the aging regime FDT can be shown to be replaced by a
 modified or generalized form commonly referred to as GFDT, and
 involves a parameter $x_c$ similar to Parisi's parameter for 1-step
 RSB. Only 1-step RSB is necessary for the case of short range
 correlations.( In the long range case that was discussed in 
Ref.~[\onlinecite{gold2}] in the equivalent context of a quantum 
particle in a random potential full RSB applies).  

The stationary regime represents the dynamics of a chain trapped in a
local minima of the random potential. Eventually after very long time
the chain can escape from its pinning and hop to another minimum
elsewhere leading to history dependence and violation of equilibrium theorems. 
In the long time limit, for a short ranged correlated random potential, the 
dynamical equations become identical to the equations of the statics
as derived from the replica method except for one equation that
involves $x_c$, which is different from the equation derived in the
statics using replicas.
This is probably due to the fact that starting from random
initial conditions the dynamics gets influenced by large barriers
and does not explore the potential landscape as efficiently as to
reproduce the statics even at large times. Thus the subtle $\ln {\cal V}$
dependence of the statics that emerges from an Imry-Ma type argument
even for the case of a zero-dimensional object in a random potential
is not reproduced by the dynamical equations.
Our results are based of course on the Gaussian variational
approximation but the emerging picture is probably valid. 
 
\acknowledgments
This research was supported in part by the US Department of Energy (DOE)
under grant No. DE-FG02-98ER45686. It was finished while I was
visiting the Weizmann Institute and I thank the Weston Visiting
Professors program for support. I also thank Leticia Cugliandolo for
a useful discussion.
 
\appendix
\section{Random potential with quadratic correlations} 
 In this Appendix we discuss the exactly solvable case of a potential
 with long range quadratic correlations \cite{shifgold1}. Since the
 variational approximation becomes exact for such a potential, our
 dynamical equations should reproduce the solution found using
 the statics and replica formalism. For this case $J(a)=(1/2)g
 a+ \rm const$, 
 and so is $\hat J(a)$ since $\gamma=0$. Thus $\hat J'(z)= g/2$, a
 constant. (The notation here is slightly different from
 Ref.~[\onlinecite{shifgold1}] where we used $J(a)=2\sigma a +\rm const$, so
 $g\rightarrow 4\sigma$.) Only the ergodic
 case applies in this case since $E_0=E_{0,0}$. We find
 \begin{eqnarray}
 E_0=\frac{gL}{T},\\
 F_k= \frac{gL}{T}(1-\delta_{k,0}). 
 \end{eqnarray}
We also obtain
\begin{eqnarray}
 \tilde q_k=\frac{gL}{\mu^2}\ \delta_{k,0}+\frac{T}{\mu+Mk^2},\\
 q_k=\frac{gL}{\mu^2}\delta_{k,0}. 
\end{eqnarray}
From these results it follows that
\begin{eqnarray}
\tilde q= \frac{g}{\mu^2}+\frac{T}{L}\sum_k
  \frac{1}{Mk^2+\mu} \rightarrow
 \frac{g}{\mu^2}+\frac{T}{2\sqrt{M\mu}}, 
\end{eqnarray}
 where the last expression applies for large $L$, and also
 \begin{eqnarray}
  q=\frac{g}{\mu^2}. 
 \end{eqnarray}
The correlation function $\tilde b(L)$ is given by
\begin{eqnarray}
 \tilde b(L)=\frac{1-e^{-L\sqrt{\mu/M}}}{\sqrt{M\mu}}. 
\end{eqnarray}
For small $\mu$ this function becomes equal to $L/M$ as in the free
case. These results coincide with the results obtained in
Ref.~[\onlinecite{shifgold1}].

Actually in this case one can write a closed form solution for
Eqs.(\ref{eq:emC})-(\ref{eq:emR}). The solution is
\begin{eqnarray}
  C_k(t.t')=\frac{gL}{\mu^2}\delta_{k,0}\left(1-e^{-\mu t}-e^{-\mu t'}+e^{-\mu(t+t')}\right)+\nonumber\\
\left(C_k(0,0)-\frac{T}{\mu+Mk^2}\right)\ e^{-(\mu+Mk^2)(t+t')}+\frac{T}{\mu+Mk^2}\ e^{-(\mu+Mk^2)|t-t'|},\\
R_k(t,t')=\theta(t-t')\ e^{-(\mu+Mk^2)(t-t')}.
\end{eqnarray}
Indeed for large times $C(t-t')$ becomes TTI and depends only on the
difference $t-t'$, and the FDT holds.

\end{document}